\newcount\subeqnno
\def\bse{\begingroup
\refstepcounter{equation}
\subeqnno=\arabic{equation}
\setcounter{equation}{0}
\def\theequation{\the\subeqnno\alph{equation}}}
\def\ese{\setcounter{equation}{\the\subeqnno}\endgroup}

\newcommand{\mb}[1]{\ifmmode#1\else\mbox{$#1$}\fi}

\newcommand\al{\mb{\alpha}}

\newcommand\be{\mb{\beta}}
\newcommand\ga{\mb{\gamma}}
\newcommand\de{\mb{\delta}}

\newcommand\la{\mb{\lambda}}
\newcommand\si{\mb{\sigma}}

\newcommand\ph{\mb{\phi}}

\newcommand\Ph{\mb{\Phi}}
\newcommand\Ps{\mb{\Psi}}

\newcommand\calA{\mb{{\cal A}}}
\newcommand\calB{\mb{{\cal B}}}
\newcommand\calC{\mb{{\cal C}}}

\newcommand\calG{\mb{{\cal G}}}
\newcommand\calH{\mb{{\cal H}}}
\newcommand\calI{\mb{{\cal I}}}
\newcommand\calJ{\mb{{\cal J}}}
\newcommand\calK{\mb{{\cal K}}}
\newcommand\calL{\mb{{\cal L}}}
\newcommand\calM{\mb{{\cal M}}}
\newcommand\calN{\mb{{\cal N}}}

\newcommand\calP{\mb{{\cal P}}}

\newcommand\calT{\mb{{\cal T}}}

\newcommand\calV{\mb{{\cal V}}}

\newcommand{\beq}{\begin{equation}}
\newcommand{\eeq}{\end{equation}}
\newcommand{\nn}{\nonumber}
\newcommand{\bea}{\begin{eqnarray}}
\newcommand{\eea}{\end{eqnarray}}
\newcommand{\norm}[1]{\parallel \! {#1} \! \parallel}

\newcommand{\inprod}[2]{\langle {#1}, {#2} \rangle}
\newcommand{\inproda}[2]{\{ {#1}, {#2} \}}

\newcommand{\pderiv}[2]{\frac{\partial {#1}}{\partial {#2}}}
\newcommand{\x}{\mb{\times}}

\newcommand{\emb}{\mb{{\rm emb}}}
\newcommand{\Ad}{\mb{{\rm Ad}}}
\newcommand{\ad}{\mb{{\rm ad}}}
\newcommand{\tr}{\mb{{\rm tr}}}

\newcommand{\m}{{\calM}}

\newcommand{\ul}[1]{{\underline{#1}}} 

\documentstyle[12pt]{article}
\topmargin = - 0.5 cm
\begin{document}
\bibliographystyle{unsrt}

%the text input

%%%%%%%%%%%%       abstract and title page    %%%%%%%%%%%%%%%%%%%%%%

\begin{flushright}
Imperial/TP/97-98/48
\end{flushright}
\begin{center}
\LARGE{Classifying Vortex Solutions to Gauge Theories} \\
\vspace*{0.5cm}
\large{
Nathan\  F.\  Lepora
\footnote{e-mail: N.F.Lepora@ic.ac.uk}$^{,1,2}$
and T. W. B. Kibble
\footnote{e-mail: T.Kibble@ic.ac.uk}$^{,1}$\\}
\vspace*{0.2cm}
{\small\em 1) Imperial College, Theoretical Physics,\\
Prince Consort Road, London, SW7 2BZ, England\\ }
\vspace*{0.2cm}
{\small\em 2) King's College, Cambridge University\\
Cambridge, CB2 1ST, England\\} 
\vspace*{0.2cm}

{May 1998}
\end{center}

\begin{abstract}
We classify the spectrum, family structure and stability of
Nielsen-Olesen vortices 
embedded in a larger gauge group when the vacuum manifold is related
to a symmetric space.
\end{abstract}

\thispagestyle{empty}
\newpage
\setcounter{page}{1}

%theorem structures
\newcommand{\proof}[1]{\smallskip \noindent {\bf Proof} \ \ {#1} \ \
  $\Box$ \medskip}
\newcommand{\result}[2]{\medskip \noindent {\bf {#1}}{\it
\ {#2}} \medskip} 

%%%%%%%%%%%%%%%%%%%%%%%%%%%%% section 1        %%%%%%%%%%%%%%%%%%%

\section{Introduction}

The aim of this paper is to classify the structure of the solution set
of embedded Nielsen-Olesen vortices for general gauge theories. To do
this requires the concepts of embedded defects, first formally
introduced in \cite{Vach94}. In many ways this work improves on and
extends the initial treatment of vortex classification attempted in
\cite{me1}, and the companion paper to that \cite{me2} contains many
illustrative examples of the formalism discussed herewithin.

This first section is devoted to a quick review of gauge theories
coupled to a scalar field sector. This will serve to set the scene and 
establish the notation for the later sections of this paper where we
shall classify vortex solutions. 
We emphasize in particular how the symmetry breaking relates to the group
structure of gauge theory.

Fermions will be excluded from the discussion, so 
we shall only have to consider the effects of scalar-gauge
interaction. This is appropriate because 
fermions seem to merely modify the form of solution, introducing
fermionic zero modes {\em around} the background scalar-gauge
configuration, whereas here we are interested in the
background configuration itself, which is determined by the {\em 
global} topographical features of the gauge theory and its symmetry
breaking. 

\subsection{Yang-Mills Theories Coupled to a Scalar Field}

We are concerned with field theories whose basic dynamical variables
are gauge potentials $A_\mu$ and scalar fields $\Phi$.
Interaction of the scalar field $\Phi$ with the gauge potential
$A_\mu$ is specified by the gauge symmetry group $G$, a compact Lie
group acting upon the scalar field via the representation $D$ of
$G$. We consider theories of the form described by the
Lagrangian~(\ref{lag}) below. Once the gauge group and its action upon
the scalar field are specified the theory is completely determined
up to the strength of scalar field self couplings and the gauge 
coupling constants.

The gauge potential $A_\mu$ lies in the Lie algebra of $G$, which we
denote by $\calG$. The field tensor of the gauge potential is
\beq
F_{\mu \nu} := \partial_\mu A_\nu - \partial_\nu A_\mu + [A_\mu,
A_\nu] \in \calG. 
\eeq 
The local actions of the gauge group upon the scalar field and gauge
potential take the form (for $g(x) \in G$) 
\bse
\bea
\label{eq-gauget-a}
\Phi(x) \mapsto \Phi'(x) &=& D(g(x)) \Phi(x), \\
\label{eq-gauget-b}
A_\mu(x) \mapsto {A'_\mu}(x) &=& g(x) A_\mu(x) g(x)^{-1}
- (\partial_\mu g(x)) g(x)^{-1}.
\eea
\ese
The field tensor transforms under a similarity transformation
\beq
F_{\mu \nu}(x) \mapsto F_{\mu \nu}'(x) = g(x) F_{\mu \nu}(x)
g(x)^{-1}. 
\eeq
The covariant derivative of $\Phi$,
\beq
\label{cd}
D_{\mu} \Phi = \left( \partial_\mu  + d(A_\mu) \right) \Phi,\\
\eeq
transforms according to
\beq
D_{\mu} \Phi(x) \mapsto (D_{\mu} \Phi)'(x) = D(g(x)) D_{\mu} \Phi(x).
\eeq
Here $d(X)$ is the derived representation of $D$, describing
how $\calG$ acts on $\Phi$ by the relation $D(e^X) = e^{d(X)}$.  

We need inner products on $\calG$ and on the
space $\calV$ of $\Phi$ values, both of which\footnote{Note that we are
using the same symbol for inner products on
$\calG$ and $\calV$; we hope it should be clear from
the context which inner product we are considering.} 
must be invariant under the actions of  
the group $G$: on $\calV$
$\inprod{D(g)\Phi}{D(g)\Phi}=\inprod{\Phi}{\Phi}$, while on $\calG$,
$\inprod{\Ad(g)X}{\Ad(g)X}=\inprod{X}{X}$, where $\Ad(.)$ is the adjoint
action of $G$ on $\calG$ defined by $ge^Xg^{-1}=e^{\Ad(g)X}$. Suitable
forms and properties of these are defined in Appendix B.

Assembling the above, the minimal gauge-invariant Lagrangian
describing the gauge picture interaction of a scalar field with a
gauge potential is: 
\beq
\label{lag}
\calL[\Phi, A_\mu] = -\frac{1}{4} \inprod{F_{\mu \nu}}{F^{\mu \nu}}
+ \frac{1}{2} \inprod{D_\mu \Phi}{D^\mu \Phi} - V[\Phi].
\eeq
Here $V[\phi]$ is the scalar potential, describing self interaction of
the  scalar field, and is constrained to be invariant under local
(gauge) transformations\footnote{Sensible quantisation also requires
the potential to be a fourth order (or less) polynomial.}.

The gauge coupling constants manifest themselves in an interesting way.
The Lie algebra $\calG$ has a natural decomposition into
commuting subalgebras 
\beq
\label{ideals}
\calG = \calG_1 \oplus \cdots \oplus \calG_n,
\eeq
with each $\calG_f$ either simple or one-dimensional. Thus an $\Ad(G)$
invariant inner product on $\calG$ has $n$ scales, related
to the norm on each $\calG_f$. With respect to this inner product the
unit norm generators are written $q_f X_f$; these $q_f$'s are
interpreted as the gauge coupling constants appertaining to $\calG_f$,
as explicitly illustrated in sec.~(\ref{Weinberg-Salam}).  

Furthermore, taking the limit $q_f \rightarrow 0$ makes the symmetry
$\calG_f$ global. For a unit norm generator $q_f X_f$, one sees
that the effect of taking $q_f \rightarrow 0$ is to decouple the gauge
field in Eq.~(\ref{cd}), rendering $\calG_f$ global.

\subsection{Symmetry Breaking}

The form of the scalar potential $V$ may cause the fully symmetric theory
to be unstable.

The trivial background vacuum takes field values $\Phi(x)=0$ and
$A_\mu(x)=0$, and a perturbative quantum field theory around it has full
gauge symmetry $G$. However, this may {\em not} be the global energy
minimum among all background field configurations. In such a situation
the unstable background will decay to the stable background vacuum, where
$\Phi(x)=\Phi_0\ne 0$. Such a background does not respect the full
symmetries of the original theory, but only a compact subgroup $H$. In
this case, the theory may admit solitonic scalar-gauge configurations
that asymptotically tend to the stable background. In this paper we
are primarily interested in the spectrum and classification of these
configurations.

It is the form of the scalar potential $V[\Phi]$ that determines
whether, and how, the gauge symmetry is spontaneously or dynamically
broken. The stable background vacuum is the one in which $\Phi(x) =
\Phi_0$, $A_\mu(x) = 0$, where $\Phi_0$ is a {\em global} minimum of the
scalar potential. Then the quantum field theory with this vacuum is
obtained by perturbing around $\Phi_0$, and is described by the Lagrangian
\beq
\calL_H [\Psi, A_\mu] = \calL_G[\Phi_0 + \Psi, A_\mu],
\eeq
which has residual gauge symmetry
\beq
H = \{h \in G : D(h) \Phi_0 = \Phi_0\}.
\eeq
The quantum field theory contains a spectrum of massive
gauge bosons corresponding to those symmetries that are broken; these
gauge bosons have internal directions in $\m$, where
\beq
\calG = \calH \oplus \m,
\eeq
the direct sum being defined by the inner product
$\inprod{\cdot}{\cdot}$.  

Because the theory has gauge symmetry $G$ (about the trivial vacuum),
the global minimum of the scalar potential generally has degenerate
values, forming a compact manifold. This is the vacuum manifold, 
\beq
\label{vacuum manifold}
M = D(G) \Phi_0 \cong G/H;
\eeq
it is the shape of this manifold that determines the non-perturbative,
solitonic spectrum of the scalar-gauge theory. The coset space $G/H$
is homogenous, and at the identity element has tangent space $\m$. The
group $H$ acts as rotations on points of $G/H$ around the identity
element. This picture is echoed in the space $D(G) \Phi_0 \subset \calV$,
where the tangent space at $\Phi_0$ is $d(\m)\Phi_0$ and $D(H)$ acts as
rotations of points of $D(G)\Phi_0$ around $\Phi_0$. One thus interprets
$\m$ as vectors that {\em move} $\Phi_0$, and $H$ as actions that rotate
{\em around} $\Phi_0$.

The set of all symmetries of the vacuum manifold form the isometry
group  
\beq
I=\{a \in {\rm aut}(\calV) : a M = M\},
\eeq
which is generally larger than $G$. This has a corresponding induced
representation $\tilde{D}$ of $I$, reducing to $D$ of $G$, and a
subgroup 
\beq
J=\{j \in I : {\tilde D}(j) \Phi_0 = \Phi_0\},
\eeq
containing $H$. The vacuum manifold may then be re-expressed
\beq
M \cong \frac{G}{H} \cong \frac{I}{J}.
\eeq

In some cases these additional elements of $I$ represent {\em
hidden} global symmetries of the theory; we discuss the $SU(2)
\rightarrow 1$ example in Sec.~(\ref{sec-su2}). In other cases,
however, the additional elements are {\em not} symmetries of the
theory. An example is the electroweak model, discussed in
Sec.~(\ref{Weinberg-Salam}), where $I=SU(2) \x SU(2)$ is {\em
not} a symmetry of the model unless ${\sin} \theta_w = 0$. However $M$
is isomorphic to the symmetric space $I/J \cong S^3$.

In what follows we shall assume
that all hidden symmetries have been included in the symmetry group
$G$. 

\section{Vortices}
\label{sec-2}

This is the pivotal section of the paper: here we describe how
vortices are classified; whereas the rest of the paper will
be used to explain and establish these results.

The tactic is to consider the archetypal Abelian-Higgs vortex, and
then to {\em embed} it into a larger theory --- one may then examine
the associated gauge freedom by moving it continuously around within the
larger theory. This will naturally partition the set of all vortex
solutions into {\em families} of gauge equivalent vortices. We will find
that this process is equivalent to a partitioning of the tangent space
to the vacuum manifold,
$\m$, in a very natural way; this equivalence being
provided by a natural association between vortices and vectors in
$\m$. Thus by explicitly partitioning $\m$ into its constituent parts a 
classification of vortices is achieved.

Our classification of vortices is on a level separate to the question of
stability. Classification relies on the group {\em actions} on the
tangent space, whereas stability relies on the {\em
topology}\footnote{Semi-local, or dynamical, stability is slightly more
subtle; we discuss this later} of the vacuum manifold. 

\subsection{The Abelian-Higgs Vortex}

This is the archetypal model of a vortex, the Nielsen-Olesen vortex,
within the simplest theory that contains such vortices.

For simplicity, we consider only straight, static vortices, and impose
cylindrical symmetry.

The Abelian-Higgs model consists of a one-dimensional complex scalar
field coupled to a $U(1)$ gauge potential $A_\mu$, with a symmetry
breaking Landau potential. (Note that, to conform with our general
notation, we take the gauge potential $A_\mu$ to be pure imaginary.) The
Lagrangian is of the form
\beq 
\calL[\Phi, A_\mu] = -\frac{1}{4} F_{\mu \nu}^\star F^{\mu \nu}
+ \frac{1}{2} D_\mu \Phi^\star D^\mu \Phi - V[\Phi],
\eeq
with
\bse
\bea
F_{\mu \nu} &=& \partial_\mu A_\nu - \partial_\nu A_\mu,\\
D_\mu \Phi &=& \left( \partial_\mu  + q A_\mu \right) \Phi,\\
V[\Phi] &=& \frac{1}{2}\la(\Phi^\star \Phi - \eta^2)^2.
\eea
\ese
Here $\la > 0$ and $\eta$ are parameters of the potential, and $q$ the
gauge coupling constant. For $\eta^2 >0$ the trivial vacuum is
unstable and decays to a vacuum of the form $\Phi_0 = \eta
e^{i\chi}$. Perturbations around this comprise a field theory
consisting of a massive gauge field interacting with a massive scalar
field; this theory having no manifest gauge symmetry. All this
information is encoded within the symmetry breaking
\beq
U(1) \rightarrow 1.
\eeq

The gauge symmetry breaking leads to the existence of vortex
solutions.  In the temporal gauge $A_0 =0$, they have the form
of the Nielsen-Olesen Ansatz: 
\bse
\bea
\label{eq-NOa}
\Phi(r, \theta) &=& \eta f_{\rm NO}(r) e^{in \theta},\\
\label{eq-NOb}
\ul{A}(r,\theta) &=& \frac{g_{\rm NO}(r)}{r} \left( \frac{in}{q}
\right) \ul{\hat{\theta}},
\eea
\ese
with boundary conditions
\beq
f_{\rm NO}(0)=g_{\rm NO}(0)=0
, \ \ {\rm and}\ \ f_{\rm NO}(\infty)=g_{\rm NO}(\infty) =1,
\eeq
Asymptotically the configurations wind around the possible degenerate
vacua, and then over the intervening space the fields continuously
interpolate. The $n= \pm 1$ solutions are stable because there is no
way they may be {\em continuously} deformed to the vacuum
$\Phi(x)=\Phi_0$ whilst keeping the asymptotic field within the vacuum
manifold.  The asymptotic field configuration, which forms a circle in
the space of scalar field values, winds around the spatial circle at
infinity --- since this map has topological degree it may not be
continuously deformed to other inequivalent situations.

\subsection{Vortices in General Gauge Theories}

We now embed the archetypal $U(1) \rightarrow 1$ vortex
(\ref{eq-NOa},\ref{eq-NOb}) within a larger theory:
\bea
G &\rightarrow& H\nn \\
\cup &\ & \cup       \ \ \ \ \ \ \ \ \ \ \ \ \ \ \ H \cap U(1) = 1\\
U(1) &\rightarrow& 1.\nn
\eea
The general vortex Ansatz is
\bse
\bea
\label{eq-vortex-a}
\Phi(r, \theta) &=& f_{\rm NO}(r) D(e^{X\theta})\Phi_0, \\
\label{eq-vortex-b}
\ul{A}(r,\theta) &=& \frac{g_{\rm NO}(r)}{r} X \ul{\hat{\theta}}.
\eea
\ese
where $X\in \m$ is the generator of the embedded $U(1)$.  
In this paper we shall classify vortex solutions of this form.

It should be remarked that the Ansatz
(\ref{eq-vortex-a}, \ref{eq-vortex-b}) is not the most general we
could assume. The lowest-energy configuration with these boundary
conditions may in some cases be distorted from the
pure embedded-vortex solution, though the distortions
are irrelevant for the classification.

Asymptotically the scalar field winds around a geodesic
$D(e^{X\theta})\Phi_0$ on the vacuum manifold $D(G) \Phi_0$, and
nearer the core the solution continuously interpolates over a two 
dimensional surface within $\calV$ containing this geodesic. 
There could be components of the scalar  field $\Phi$ out of this
surface that vanish at infinity but are non-zero in the core
\cite{andy97}.  However, such components do not affect the
classification of the vortex solutions, so we may ignore them here. 

The asymptotic gauge field in the minimum-energy configuration is in
the direction of $X\in \cal M$, because if we were to add any
component in $\cal H$, the effect would be to leave the scalar field 
unchanged but to add an extra term to the gauge-field
magnetic energy. Though one should note that when gauge field takes
components within more than one simple part of $G$ the various
components may have different radial dependence, and so near the
core there may be components in $\cal H$. Again, however, these
components in $\cal H$ do not affect the classification.  

\subsubsection{Vortex Generators}

We shall always consider vortices in the temporal gauge, and assume
cylindrical symmetry, {\em i.e.} their field values are cylindrically
symmetric, and stationary in time. Non-trivial gauge transformations that
respect this gauge are the {\em global} (or rigid) gauge
transformations where $g(x)$ in Eq.~(\ref{eq-gauget-a},\ref{eq-gauget-b})
is independent of $x$.

By Eq.~(\ref{vacuum manifold}) we can use the gauge freedom of $G$ to
fix $\Phi_0$.  Then vortex solutions are defined by just one variable,
the vortex generator $X\in \m$.  To classify vortex solutions, we must
answer the questions: which generators $X \in \m$ define vortex
solutions?  And which of those are gauge equivalent?  

The remaining gauge freedom corresponds to global transformations by
elements $h\in H$, the residual symmetry group.  Two vortices
described by Ans\"atze of the form (\ref{eq-vortex-a},
\ref{eq-vortex-b}) are gauge equivalent if and only if they are
related by such a transformation. Since 
$\Phi_0$ is unaltered, the only effect is to transform the generator $X
\in \m$: 
\beq
X \mapsto \Ad(h) X, \ \ \ {\rm with}\ h \in H,
\eeq
where $\Ad$ is the adjoint representation of $G$ acting on $\calG$.

Geometrically, the scalar boundary
conditions of a vortex describe a geodesic $D(e^{X\theta})\Phi_0$
on the vacuum manifold, with the gauge potential being defined from
the tangent vector to the corresponding curve $e^{X\theta}$ in $G$. Global
gauge transformations defined by elements of $H$ keep the point $\Phi_0$
fixed, and rotate the geodesic around this point. Thus vortex generators
that may be rotated into one another by $\Ad(H)$ define gauge equivalent
vortices. 

There are two constraints that the vortex Ansatz must
obey~\cite{barr92}:\\  
(i) Closure. The asymptotic scalar field forms a closed geodesic
with $D(e^{2 \pi X}) \Phi_0 = \Phi_0$.  Equivalently, $e^{2 \pi X}\in
H$.\\ 
(ii) The Ansatz is a solution to the equations of motion. An argument
of \cite{Vach94} identifies this to be when: fields in the vortex
do not induce currents perpendicular (in Lie algebra space) to it. This
can be shown to be equivalent to (see appendix A): 
\begin{quote}
\em if $X$ is a vortex generator, then for all $X^{\perp}$ such that
$\inprod{X}{X^\perp} = 0$ one has
$\inprod{d(X)\Phi_0}{d(X^\perp)\Phi_0}=0$.
\end{quote}
In other words the map $X \mapsto d(X)\Phi_0$ is {\em conformal} over
the classes of generators that define vortex solutions.

\subsubsection{Family Structure of Vortex Solutions}

Consider the action of $\Ad(H)$ on $\m$.  In general, $\m$ may be {\em
reducible}; let us write
\beq
\label{eq-decomp}
\m = \m_1 \oplus \cdots \oplus \m_n
\eeq
where $\m_i$ is (real) irreducible under the action of $\Ad(H)$. In
Appendix B we show that the (real) inner products on $\calV$ and
$\calM$ are related upon this decomposition by: 
\bea
\label{crucial}
\inprod{d(X_i)\Phi_0}{d(Y_j)\Phi_0} &=& \la_i \la_j
\inprod{X_i}{Y_j}, \ X_i \in \m_i, Y_j \in \m_j, \\
{\rm where}\  \la_i &=&
\frac{\norm{d(X_i)\Phi_0}}{\norm{X_i}}\nn, 
\eea
and $\la_i$ is constant upon its particular $\m_i$. 

Comparing Eq.~(\ref{crucial}) with condition (ii) above we see that
generally, vortex generators lie
only in the individual $\m_i$'s, but for some pairs of values of gauge
coupling constants, defined by the condition,
\beq
\la_i(q_1,\cdots ,q_n) = \frac{\norm{d(X_i)\Phi_0}}{\norm{X_i}} =
\frac{\norm{d(X_j)\Phi_0}}{\norm{X_j}} = \la_j(q_1,\cdots ,q_n),
\eeq
the
solution space increases, encompassing {\em combinations} of vortices
with generators in $\m_i$ and vortices with generators in $\m_j$.
Then the vortex generators are those generators that close in $\m_i
\oplus \m_j$; such vortices we dub {\em combination vortices}.

Leaving aside the special case of combination vortices, we may
concentrate on vortices described by generators $X$ in one specific
irreducible subspace $\m_i$. First, we show that every $X$ is
equivalent to one lying in a subspace that plays the same role as a
Cartan subalgebra in a Lie algebra. 

For each $\m_i$ one may define a {\em rank} analogous to the rank of a
Lie group: the rank $r$ of $\m_i$ is the maximum number of linearly
independent commuting generators in $\m_i$.  Let $T_1,\ldots,T_r$ be
such a set, and consider the algebra, $\calT$ say, that they generate. 
It is an Abelian subalgebra of $\calG$, and $T=\exp\calT$ is an Abelian
subgroup of $G$, a torus. 

Then, in general we conjecture that vortex generators are given by the
following crucial result:
\begin{quote}\em 
for $\m_i$ of rank $r$, we can find a set of linearly independent
commuting generators $\{T_1, \ldots, T_r\}$ such that the $X \in \m_i$
satisfying the closure condition may be written 
\beq
\label{result}
X = \Ad(h) \sum_{k=1}^{r} l_k T_k,
\eeq
with $h \in H$, and $l_k\in{\bf Z}$.
\end{quote}

In sec.~(\ref{sec-proof}) we prove the above result for two categories
of vacuum manifolds: symmetric spaces, and symmetric spaces that are 
modified to admit semi-local vortices. Practically, these two
situations cover most cases of physical relevance, although we suspect
that the result is true generally.

\subsubsection{Gauge Equivalent Embedded Vortices}

Each of the vortices labelled by the set of coefficients $\{l_j\}$ has
an orbit of gauge equivalent solutions associated with it,
\bse
\bea
\Ad(H) \sum_{k=1}^{r} l_k T_k \cong H/H_{\{ l_j\}},
\ \ \ \ \ \ \ \ \ \ \ \ \ \ \ \ \\
{\rm where}\ \ \ H_{\{ l_j\}} = \{h\in H :
\Ad(h)\sum_{k=1}^{r} l_k T_k = \sum_{k=1}^{r} l_k T_k\}.
\eea
\ese
Generally these orbits differ in structure depending upon which
element of the lattice of coefficients is considered.

In addition, these orbits may contain vortex generators
with different sets of coefficients $\{l_j\}$ that are  gauge
equivalent.  In fact, 
$H$ acts as a discrete transformation group on the lattice $\ul L$.  We
can construct this group as follows.
 
First, define the subgroup of $H$ that maps $\calT$ into itself: 
\beq
N({\calT}) = \{h\in H: \Ad(h){\calT}={\calT}\}.
\eeq
Then, introduce the subgroup which leaves each point of the lattice
unaltered:
\beq
B({\calT}) = \{h\in H: \Ad(h)T_k=T_k, k=1,\ldots,r\}.
\eeq
The effective transformation group on the lattice $\ul L$ is the
quotient group $K=N({\calT})/B({\calT})$.  Thus finally the set of
gauge-inequivalent vortex generators in $\m_i$ is the quotient ${\ul
L}/K$, comprising equivalence classes of lattice points under the
induced action of $K$.

We conjecture that the effect of $K$ upon the lattice $\ul L$ allows
us to enumerate gauge inequivalent vortex generators by $l_1 \geq
\cdots \geq l_r$ for those $\calM_i$'s of dimension larger than
one. This conjecture is largely motivated by example (\ref{sec-su5}),
and is beyond the scope of the present work to prove.

This completes the classification of vortex generators in each of the
subspaces $\m_i$.  Of course, if the coupling constants allow then one
may also have combination vortex generators, as described above.

%%%%%%%%%%%%%%%%%% proving our theorem sec 3 %%%%%%%%%%%%%%%%%%%%%%

\section{Symmetric Spaces and the Conjugacy of Maximal Tori}
\label{sec-proof}

The goal of this section is to discuss the mathematics of, and prove,
the following crucial result upon vacuum manifolds that are either
compact symmetric spaces or compact symmetric spaces that are modified
to admit semi-local vortices: 
\begin{quote}\em 
for $\m_i$ of rank $r$, we can find a set of linearly independent
commuting generators $\{T_1, \ldots, T_r\}$ such that the $X \in \m_i$
satisfying the closure condition may be written 
\beq
X = \Ad(h) \sum_{k=1}^{r} l_k T_k,\nn
\eeq
with $h \in H$, and $l_k\in{\bf Z}$.
\end{quote}

Before establishing the above result for our cases of interest, we
shall discuss some basic properties of symmetric spaces that will be
useful as a background to the work in this paper~\cite{Helg}. 

There are several ways of characterising when the vacuum manifold
$G/H$ is a symmetric space and we find the most convenient in the
context of this work to be:  
$G/H$ is a symmetric space if the decomposition $\calG = \calH
\oplus \calM$ obeys
\beq
\label{symspace}
[\calM, \calM] \subseteq \calH.
\eeq

This has the consequence (Cartan, \cite{cartd}) that $M$ may be
written as $M \cong \exp(\calM)$, and then $G$ has the polar
decomposition into cosets $G = \exp(\calM)H$. Geodesics in
$\exp(\calM)$ are easily found, and those passing through $\bf 1$ may
be written as $\ga_X(t) = \exp(Xt)$ with $X \in \calM$. 

Other important features of symmetric spaces may be found from the
above structure, such as a well defined composition of elements in $M$
relating to a symmetry operation upon geodesics. However these
features are not central to this paper and we shall not discuss them
further. 

Cartan has proved that the compact symmetric spaces can be written as
$M = M_0 \x M_+$; where $M_0$ is the toroidal part of $M$, written
$M_0=U(1)^k$, and $M_+ \cong G/H$, with $G$ semi simple, is the
rest. We shall establish the main result of this section firstly upon
$M_0$, and then $M_+$, before discussing the case when $M$ has been
modified from a symmetric space so as to admit semi-local vortices.

\subsection{Toroidal $M_0$}
\label{tor}
For the toroidal part of the vacuum $M_0$ we may write $M_0 \cong G/H$
with $G=U(1)^k$  and $H = {\bf 1}$. Trivially then $\calM_0 = \calG =
u(1)^k$ and then $[\calM_0, \calM_0] = 0 \subseteq \calH$.
Eq.~(\ref{result}) is trivial in such a case, since each $\calM_i$ is 
one-dimensional. 

\subsection{Non-toroidal $M_+$}

For the non-toroidal part of a symmetric vacuum $M_+$, Cartan has
shown that $M_+ \cong G/H$ where $G$ is a semi-simple Lie group and 
$H$ is embedded in $G$ according to Eq.~(\ref{symspace}) above. We
now establish the main result of this section for such cases.

Firstly we establish the following, a variant of Hunt's lemma
\cite{Hunt}:  
\begin{quote}\em
for $X, Y \in \m_i$ there exists $h_0 \in H$ such that $[\Ad(h_0)X, Y]
=0$.
\end{quote}
This result is proved by considering a function $f: H \to {\bf R}$
defined by
\beq
f(h) = \inprod{Ad(h)X}{Y}.
\eeq
Since this is a bounded real function on a compact domain, it attains it
maximum, say at a point $h_0$. Then for any
$K\in \calH$,
\beq
0=\left. \frac{d}{dt}f(e^{Kt}h_0)\right|_{t=0}=
\inprod{\ad(K)\Ad(h_0)X}{Y}=\inprod{K}{[\Ad(h_0)X,Y]}.
\eeq
Finally, since $G \rightarrow H$ defines a symmetric space,
$[\Ad(h_0)X, Y] \in \calH$ and therefore vanishes, proving the lemma.

From this result, in three steps we may establish our main result when
the vacuum is a symmetric space.

Firstly, there exists $Y \in
\m_i$ such that the centraliser $\calC$ of $Y$ in $\m_i$
\beq
\calC(Y) = \{ U \in \m_i : [Y,U]=0 \},
\eeq
is precisely $\calT$. This follows by considering elements $Y$ that
define paths $\{\exp(\al Y): \al \in {\bf R}\}$ dense in $T =
\exp(\calT)$. Then for any $Z \in \calM_i$ such that $[Z, Y]=0$, 
also $[Z, \calT]=0$ and thus, by maximality, $Z \in \calT$. 

Next, we use the result established above. Thus for any $X \in
\m_i$ there exists $h_0 \in H$ such that $\Ad(h_0)X \in
\calC(Y)$. Hence, we have shown that there exists $h_0 \in H$ with
$\Ad(h_0)X \in \calT$. 

Finally, we note that within $\calT$, the generators
$X$ that satisfy the condition of closure, $e^{2\pi X}\in H$, obviously
form a lattice, $\ul L$ say, and we can choose the generators $T_j$ so
that this comprises precisely the linear combinations with integer
coefficients, $\sum_j l_j T_j$, $\l_j\in{\bf Z}$.  

\subsection{$M$ Admits Semi-local Vortices}

Here we deal with the situation when $M \cong G/H$ admits semi-local
vortices, and has been modified from a symmetric space.
Specifically we are going to consider vacuum manifolds of the form
\beq
\label{nonsym}
M \cong \frac{K \x A/Z}{H}, 
\eeq
where $K$ is semi-simple, $A$ is Abelian and connected, $Z$ is a
finite subgroup, and $H$ is isomorphic to a subgroup $L$ of $K$, such
that $K/L$ is a compact symmetric space. 

Perhaps the easiest way to see the structure of Eq.~(\ref{nonsym}) is
to represent it by a two step symmetry breaking:
\bea
\calG = \calK \oplus \calA &\rightarrow& \calL \oplus \calA = \calH'
\oplus \calC(L) \oplus \calA \nn \\ 
           &\rightarrow& \ \calH \  = \calH' \oplus \calB,
\eea
where here $H'$ is the semi-simple part of $L$, $\calC(L)$ is the
connected centre of $\calL$, and $\calB \cong \calC(L)$.

Decomposing $\calG = \calH \oplus \calM$, we see that $\calM$ is
made up of two parts, and may be written as $\calM = \calP \oplus
\calB^\perp$, where
\bse
\bea
\calK &=& \calL \oplus \calP,\\
\label{calB}
\calC(L) \oplus \calA &=& \calB \oplus \calB^\perp.
\eea
\ese
From this decomposition the main result~(\ref{result}) of this section
can be established: it is trivially true upon $\calB^\perp$ and,
because $K/L$ is a symmetric space, it is true also on
$\calP$.

%%%%%%%%%%%%%%%%%%%%%%% stability sec4 %%%%%%%%%%%%%%%%%%%%%%%%%%%%

\section{The Stability of Vortices in General Gauge Theories}

The stability of the vortices is on a level separate from the family
structure discussed above, though clearly all vortices in the same
family have the same stability structure since all such vortices are
gauge equivalent. 

Classical stability is determined by the energetics of the vortex
solution. If it is energetically favourable for a vortex solution to
continuously relax to the vacuum then that vortex is unstable. This is
the general situation for vortex solutions, and generally one needs
some property of the scalar-gauge theory to `prop-up' the solution, so
as to make the configuration a local {\em minimum} of the energy.
There are two ways this may be achieved: either through the topology
of the vacuum manifold, or through the dynamics of the
solution~\cite{Vach91}. We discuss each of these cases separately.

\subsection{Topological Stability}

Topological stability arises through the vacuum manifold not being
simply connected, so that
\beq
\label{eq-homotopy}
\pi_1 (G/H) \neq 0.
\eeq
Vortex families whose boundary conditions are the elements of the
non-trivial homotopy classes of Eq.~(\ref{eq-homotopy}) are imbued
with a {\em conserved} topological charge, which in appropriate cases
guarantees their stability. Physically, the vacuum manifold contains
loops that are incontractible, so that the vortex solutions corresponding 
to such loops are topologically obstructed from continuously deforming
to the vacuum.

There are two distinct ways in which topological stability may occur:
either through an Abelian or a non-Abelian part of the symmetry breaking.
These two different situations also have contrasting properties for their
family structure, and the nature of their stability. Therefore we discuss
these two cases separately.

Abelian topological vortices are present for a symmetry breaking scheme
such as
\beq
G = G' \x A/Z \rightarrow H \subseteq G',
\eeq
where $A$ is an Abelian subgroup, a torus
\beq
A= U(1)_1 \x \cdots \x U(1)_N
\eeq
and $Z$ is a finite group. Then $\pi_1(G/H)$ takes the form  
\beq
\pi_1 (G/H) = P \x{\bf Z}^N,
\eeq
where $P$ is a finite group. The Abelian topologically stable vortices
have boundary conditions that belong to non-trivial ${\bf Z}^N$ homotopy
classes.

The Lie algebra of $G$ decomposes under the adjoint action of $H$ into
\beq
\calG = \calH \oplus \m' \oplus \calA.
\eeq
The Abelian subalgebra $\calA$ is generated by $N$ commuting generators
$\{T_k\}$.  We can always choose these so that the elements satisfying the
closure condition are given by $\sum_k l_k T_k$, though only some of these
may actually correspond to stable vortices.  Vortices with distinct
values of the quantum numbers $l_k$ are necessarily gauge inequivalent;
there is no non-trivial family structure.

Note that $\calA$ may be described as the intersection of $\m$ and the centre
$\calC$ of the Lie algebra $\calG$.

Non-Abelian topological stability can arise from elements outside the
centre of the group; if for example the vacuum manifold $G/H$ has the form
of the quotient of a simply connected manifold $M$ by a discrete
group ${\bf Z}_n$, then $\pi_1(G/H) = \pi_1(M/{\bf Z}_n) = {\bf Z}_n$.
This gives the vortex generators in the relevant $\m_i$ subspaces a
mod-$n$ topological charge, imparting stability to the corresponding
vortex solutions. Such stable vortices correspond to {\em fractional
quantum} embedded vortices, and the solutions have non-trivial family
structure because necessarily ${\rm dim}(\m_i)>1$.

\subsection{Dynamical Stability}
\label{dynamical}

Even when there is no topology to stabilise a vortex solution
there may be non-trivial dynamics making the decay modes of
the vortex energetically unfavourable: this is called dynamical
stability~\cite{Vach91}. Often this occurs when an Abelian symmetry
couples much more strongly than a non-Abelian symmetry; then the
decay modes cost large gradient energies that cannot be
compensated by inducing a non-Abelian gauge field.

Such vortices are solutions to gauge theories that have gauge coupling 
constants close to a `semi-local limit', where the scalar-gauge theory
admits semi-local vortices. Preskill \cite{Pres92}, has shown that
semi-local vortices are solutions to gauge theories of the form
\beq
G_{\rm global} \x U(1)_{\rm local} \rightarrow H, \ \ \ {\rm with}\ 
H \cap U(1)_{\rm local}={\bf 1},
\label{eq-22}
\eeq
where the suffices `global' and `local' represent non-gauge and gauged
symmetries, respectively. Requiring that this does not admit
topological vortices leads to the condition $H \not\subset
G_{\rm global}$. Note that Eq.~(\ref{eq-22}) need only be a sub-part
of a more general symmetry breaking. 

Thus we define a `semi-local limit' of a gauge theory to be a limit of
the gauge coupling constants $\{ q_f \}$ in which part of the gauge
theory takes the form in Eq.~(\ref{eq-22}). Recall that the formal
limit of taking a coupling constant 
$q_f \rightarrow 0$ decouples the gauge field and 
makes the corresponding gauge symmetry global. Thus dynamically stable
vortices are generated by the corresponding generator of semi-local
vortices, but with the gauge coupling constants {\em close} to those
in the semi-local limit.

The above may be expressed succinctly in terms of the group theory,
describing which vortex generators give solutions that may be
dynamically stable.  Recall that $\calC$ is the centre of the Lie
algebra $\calG$, and denote the projection of an element $X \in \calG$
into $\calM$ by ${\rm pr}_\calM(X)$.  We then have the following:   
\begin{quote}\em
vortex generators $X \in {\rm pr}_\calM(\calC)$ such that $X \not\in
\calC$ define embedded vortices that 
are stable in a well defined semi-local limit of the model and are
thus dynamically stable for a region of parameter space around that
semi-local limit. 
\end{quote}
We also have a useful subresult about the dimension of $\calM_i$'s for
such vortices:
\begin{quote}\em
vortices with a stable semi-local limit are {\em always} generated by
generators in one dimensional $\calM_i$'s.
\end{quote}
The proofs are given in Appendix C.

Together these give a nice interpretation of the role of semi-locality:
considering $\calM_i \subset {\rm pr}_\calM(\calC)$, then the
embedding $\calM_i \subset \calG$ is determined by the coupling
constants $\{ q_i\}$. Essentially the semi-local limit is for
coupling constants such that the embedding is $\calM_i \subset
u(1)_{\rm local}$. Then dynamical stability occurs when the embedding
is close to this.

\section{Examples}

This section illustrates principles discussed generally in the
preceding sections. 

\subsection{$SU(2) \rightarrow 1$}
\label{sec-su2}

This is an examples of a symmetry breaking that contains hidden
symmetries. 

Clearly, $\calH=0$ and $\calM= su(2)$. Then, since $\calM$ is a Lie
algebra, the relation $[\calM, \calM] = \calM$ holds. 

However the vacuum manifold is a three-sphere embedded in ${\bf C}^2$,
with an isometry group
\beq
I = I(S^3) = SU(2)_l \x SU(2)_g,
\eeq
which is larger than the gauge group. The $SU(2)_l$ represents the
left actions upon a complex doublet, with generators
represented by $X_a = -\frac{1}{2}i \si_a$, and the $SU(2)_g$ is a
hidden global symmetry with generators $Y_1= -\frac{1}{2} \si_2 K, \ 
Y_2=\frac{1}{2} i \si_2 K$, and $Y_3= -\frac{1}{2}i {\bf 1}_2$. Here
$K$ is the complex conjugation operator and $\si_a$ are the Pauli spin
matrices.

Then the stability group of $I$ upon a point of $M$ is
\beq
J = SU(2)_{l+g},
\eeq
the diagonal subgroup lying {\em between} $SU(2)_l$ and $SU(2)_g$. 
Hence the actual symmetry breaking is
\beq
\label{eq-fullsb}
SU(2)_l \x SU(2)_g \rightarrow SU(2)_{l+g},
\eeq
which contains the gauge symmetry breaking $SU(2)_l \rightarrow 1$,
and has additional global symmetry $SU(2)_g$. Defining $\calI = \calJ
\oplus \calN$ one can easily verify $[\calN, \calN] \subseteq \calJ$.

It is interesting to note that the above model is the Weinberg-Salam
model, with electroweak mixing angle $\sin{\theta_w}=0$.

\subsection{$SU(2) \x U(1) \rightarrow U(1)$}
\label{Weinberg-Salam}

The gauge symmetry $SU(2) \x U(1)$ acts on a two-dimensional complex
scalar field $\Ph$ by the fundamental representation. The generators
of $SU(2)$ 
are $X_a = -\frac{1}{2}i \si_a$ and the $U(1)$ generator is  
$X_0 =\frac{1}{2} i {\bf 1}_2$. 

The vacuum manifold is 
\beq
M \cong \frac{SU(2) \x U(1)}{U(1)} \cong \frac{SU(2) \x 
SU(2)}{SU(2)_{\rm diag}},
\eeq
thus the isometry group is $I= SU(2) \x SU(2)$. This is in general
not a symmetry group of the theory unless ${\sin}\theta_w=0$.

The inner product on $su(2) \oplus u(1) \subset gl({\bf C}^2)$ is
obtained from Appendix B:
\beq
\inprod{X}{Y} = -2 s_1 \tr XY - \frac{s_2}{4}\tr X \tr Y.
\eeq
To obtain the usual scalar-gauge coupling~(\ref{couples}), $s_1$ and
$s_2$ are related to the hypercharge $g'$ and isospin $g$ coupling
constants such that  
\beq
\inprod{X}{Y} = -\frac{1}{g^2} \left\{ 2{\tr}XY + (\cot^2 \theta_w-1)
{\tr} X {\tr} Y \right\},
\eeq
with $\tan \theta_w = g'/g$.
The unit norm generators are then $gX_a$ and $g'X_0$, and thus the
components of a gauge field $A^\mu$ couple to $\Ph$ as
\beq
\label{couples}
A^\mu \Phi = (gW^\mu_a X_a + g' Y^\mu X_0) \Phi.
\eeq	

Taking the vacuum to be defined from $\Ph_0 = v (0\ 1)^\top$ the
decomposition $\calG = \calH \oplus \calM$ takes the form
\beq
\calH = \left( \begin{array}{cc} i \al & 0 \\ 0 & 0 \end{array}
\right)\ \ \ \ {\rm and}\ \ \ \ 
\m = \left( \begin{array}{cc} -i\be \cos 2\theta_w & \ga \\ -\ga^* & 
i\be \end{array} \right), 
\eeq
with $\al, \be$ real and $\ga$ complex. Then under $\Ad(H)$,
$\calM$ decomposes to the irreducible subspaces
$\calM = \calM_1 \oplus \calM_2$, with
\beq
\calM_1 =
\left( \begin{array}{cc} -i\be \cos 2 \theta_w & 0 \\ 0 & i \be
\end{array} \right) \ \ \ \ {\rm and}\ \ \ \ 
\calM_2 =
\left( \begin{array}{cc} 0 & \ga \\ -\ga^* & 0 \end{array}
\right). 
\eeq
Additionally ${\rm pr}(u(1)_Y) = \calM_1$.

Thus we obtain a one-parameter family of gauge equivalent unstable
embedded vortices and a gauge invariant embedded vortex with a stable
semi-local limit. These are identified as the W-strings and Z-string
of the Weinberg-Salam model.

\subsection{$SU(5) \rightarrow SU(3) \x SU(2) \x U(1)$}
\label{sec-su5}

This is an example of a symmetry breaking where the irreducible
$\m_i$'s are of a non-trivial rank.
 
The gauge group $G=SU(5)$ acts on the 
scalar field $\Ph \in \calG$ by the adjoint action. For vacuum
expectation value
\beq
\Ph_0 = iv \left( \begin{array}{ccc} \frac{2}{3} \bf{1}_3 & \vdots &
\bf{0} \\ \cdots & \cdots  & \cdots \\ \bf{0} & \vdots & -\bf{1}_2
\end{array} \right), 
\eeq
$G$ breaks to $H = SU(3)_c \x SU(2)_I \x U(1)_Y$:
\beq
\left( \begin{array}{ccc} SU(3)_c & \vdots &
\bf{0} \\ \cdots & \cdots  & \cdots \\ \bf{0} & \vdots & SU(2)_I
\end{array} \right)
\x
\left( \begin{array}{ccc} e^{\frac{2}{3}i \theta} {\bf 1_3} & \vdots &
\bf{0} \\ \cdots & \cdots  & \cdots \\ \bf{0} & \vdots & 
e^{-i \theta}{\bf 1}_2 \end{array} \right)
\subset SU(5).
\eeq

Reducing $\calG$ into $\calG = \calH \oplus \m$, where
\beq
\m = \left( \begin{array}{ccc} \bf{0}_3 & \vdots &
\underline{A} \\ \cdots & \cdots  & \cdots \\ -\underline{A}^\dagger 
 & \vdots & \bf{0}_2
\end{array} \right),
\eeq
one finds this to be irreducible under the adjoint action of $H$. 
However it is clear that ${\rm rank}(\m)=2$, for example with
generators $T_1, T_2$ such that
\beq
T_i = \frac{1}{2}\left( \begin{array}{ccc} \bf{0}_3 & \vdots &
\underline{A}_i \\ \cdots & \cdots  & \cdots \\
-\underline{A}_i^\dagger & \vdots & \bf{0}_2
\end{array} \right), \ 
{\rm with}\ 
\underline{A}_1 = \left( \begin{array}{cc} 1 & 0 \\ 
0 & 0 \\ 0 & 0 \end{array} \right),\ \ 
\underline{A}_2 = \left( \begin{array}{cc} 0 & 0 \\ 
0 & 1 \\ 0 & 0 \end{array} \right),
\eeq
then $[T_1, T_2]=0$ and there are no other linearly independent
generators that commute with both of these.

Then using Eq.~(\ref{result}) the vortex generators in $\m$ are:
\beq
\label{eq-su5gen}
X = \Ad(h) \left(l_1 T_1 + l_2 T_2 \right),
\eeq
with $l_i \in {\bf Z}$ and $h \in H$. These generators form a ${\bf
Z}^2$ lattice of solutions $(l_1, l_2)$, with an orbit of solutions
generated by $\Ad(H)$ from each point of the lattice. However, there
is degeneracy since it is easy to find $h_1, h_2 \in H$ such that 
\bse
\bea
\Ad(h_1) (l_1 T_1 + l_2 T_2) &=& -l_2 T_1 + l_1 T_2, \\
\Ad(h_2) (l_1 T_1 + l_2 T_2) &=& -l_1 T_1 + l_2 T_2.
\eea
\ese
Together these generate the discrete group of actions $\Ad(K)$ that
takes the pair $(l_1, l_2)$ to the set of eight pairs $\{  (\pm l_1,
\pm l_2), (\mp l_1, \pm l_2)$, $(\pm l_2, \pm l_1), (\mp l_2,  \pm
l_1) \}$.   
Then the set of gauge inequivalent embedded vortex generators in $SU(5)$
are those in Eq.~(\ref{eq-su5gen}), but with $(l_1, l_2)$ restricted
to the octant $l_1 \geq l_2 \geq 0$.

To explicitly give an example of a vortex we give the $(1,0)$ solution:
\bse
\bea
\Phi(r, \theta) &=& f_{\rm NO}(r) \Ad(e^{T_1\theta})\Phi_0, \\
\ul{A}(r,\theta) &=& \frac{g_{\rm NO}(r)}{r} T_1 \ul{\hat{\theta}}.
\eea
\ese
then, with a little calculation
\beq
\Ad(e^{T_1\theta})\Phi_0= \Phi_0 - \frac{5v\sin \frac{\theta}{2}
A}{3}( \sin \frac{\theta}{2} {\bf 1}_5 + \cos \frac{\theta}{2}
T_1),  
\eeq
where $A_{ij}=i(\de_{i1}\de_{j1} - \de_{i4}\de_{j4})$. A non-trivial
geodesic, that is certainly not contained within a plane through the
origin. 

\section{Conclusions}

We conclude by briefly summarising our classification.

We consider minimal, gauge invariant theories~(\ref{lag}) of symmetry
breaking $G \rightarrow H$, where $H \subset G$ are compact Lie
groups. Then the following structure is required: writing $\calG =
\calH \oplus \calM$, one splits $\calM =\calM_1 \oplus \cdots \oplus
\calM_n$, the irreducible subspaces of $\calM$ under $\Ad(H)$. 

Then cylindrically symmetric, time independent vortex solutions are of
the form  
\bea
\Phi(r, \theta) &=& f_{\rm NO}(r) D(e^{X\theta})\Phi_0, \nn \\
\ul{A}(r,\theta) &=& \frac{g_{\rm NO}(r)}{r} X \ul{\hat{\theta}}. \nn 
\eea
with $X \in \calM_i$ specified by the following result, which has been
rigorously established for $G/H$ a symmetric space or a symmetric
space modified to admit semi-local vortices: 
\begin{quote}\em
for $\m_i$ of rank $r$, we can find a set of linearly independent
commuting generators $\{T_1, \ldots, T_r\}$ such that the $X \in \m_i$
satisfying the closure condition may be written 
$$
X = \Ad(h) \sum_{k=1}^{r} l_k T_k, 
$$
with $h \in H$, and $l_k\in{\bf Z}$.
\end{quote}
 
Generally this classifies {\em all} vortices of the embedded
Nielsen-Olesen type, however for certain critical values of the
ratios of gauge coupling constants the solution set may increase
to include combination vortices also, with generators $X$ lying
between $\calM_i$ and $\calM_j$. 

Stability of vortices may be of two types. Firstly, topological
stability given by the non-trivial first homotopy classes of
$G/H$. Secondly, dynamical stability, specified by the following
result:
\begin{quote}\em
vortex generators $X \in {\rm pr}_\calM(\calC)$ such that $X \not\in
\calC$ define embedded vortices that 
are stable in a well defined semi-local limit of the model and are
thus dynamically stable for a region of parameter space around that
semi-local limit. 
\end{quote}
Such vortex generators always lie in one-dimensional $\calM_i$'s.

We conjecture that the above classification also holds true in
general. 

\bigskip

%%%%%%%%%%%%%%%%%%%%%%%%%%%%% acknowledgements %%%%%%%%%%%%%%%%%%%

{\noindent{\Large{\bf Acknowledgements.}}}
\nopagebreak
\bigskip
\nopagebreak

This work was supported in part by PPARC.  N.L. acknowledges
EPSRC for a research studentship and thanks A.C. Davis for
interesting discussions related to this work. This work was supported
in part by the European Commision under the Human Capital and Mobility
program, contract no. CHRX-CT94-0423.

\bigskip
\bigskip

%%%%%%%%%%%%%%%%%%%%%%%%% appendix  %%%%%%%%%%%%%%%%%%%%%%%%%

{\noindent{\Large{\bf Appendix A}}} 
\par
\nopagebreak
\bigskip
\nopagebreak
\noindent
The Lagrangian (\ref{lag}) gives field equations
\bse
\bea
D^\mu D_\mu \Phi &=& -\pderiv{V}{\Phi}, \\
\inprod{D^\mu F_{\mu\nu}}{X} &=& \inprod{J_\nu}{X} =
\inprod{d(X)\Phi}{D_\nu \Phi} - \inprod{D_\nu \Phi}{d(X)\Phi}.
\eea
\ese

The vortex Ansatz
\bse
\bea
\Phi_{\rm vor}(r, \theta) &=& f_{\rm NO}(r) D(e^{X\theta})\Phi_0, \\  
\ul{A}_{\rm vor}(r,\theta) &=& \frac{g_{\rm NO}(r)}{r} X
\ul{\hat{\theta}}, 
\eea 
\ese
naturally splits $\calG$ globally into 
$\calG = \calG_\emb \oplus \calG_\emb^\perp$ such that 
$\inprod{\ul{A}_{\rm vor}(x)}{\calG_\emb} \neq 0$ for all nonzero $x$.
In the scalar sector it provides a local decomposition
$\calV = \calV_\emb(\theta) \oplus \calV_\emb^\perp(\theta)$ where 
$\calV_\emb(\theta) = D(e^{X \theta})({\bf R}\Phi_0 + {\bf R} 
d(X_\emb) \Phi_0)$, with ${\bf R}$ the real numberline.
 
Substituting this vortex Ansatz into the field equations and requiring
it to be a solution yields
\bse
\bea
\inprod{D^\mu D_\mu \Phi_{\rm vor}(x)}{\calV_\emb^\perp(\theta)} &=&
0,\\ 
\inprod{J^\mu(x)}{\calG_\emb^\perp} &=& 0,
\eea
\ese
from which one obtains,
\bse
\bea
\inprod{\Psi}{\pderiv{V}{\Phi}{[\phi]}}  
&=& 0,\ \  \Ps \in \calV_\emb^\perp(\theta), \ph \in
\calV_\emb(\theta), \\ 
\langle d(X^\perp)\Ph, \calV_\emb(\theta) \rangle &=& 0,\ \ 
\Ph \in \calV_\emb(\theta), X^\perp \in \calG_\emb^\perp.
\eea
\ese
The first condition constrains the scalar potential and may restrict
combination vortex solutions. The second we rephrase:
using the identity $\inprod{d(\calG)\Ph}{\Ph}=0$ for $\Ph \in \calV$, 
\bea
\inprod{d(X^\perp)\Ph}{\calV_\emb(\theta)} =0 
&\Leftrightarrow&
\inprod{d(X^\perp) D(e^{X\theta})\Ph_0}{D(e^{X \theta})d(X)\Phi_0}=0
\nn \\  
&\Leftrightarrow&
\inprod{d(\Ad(e^{-X \theta})X^\perp)\Ph_0}{d(X)\Ph_0} = 0, \\
&\Leftrightarrow&
\inprod{d(X^\perp)\Ph_0}{d(X)\Ph_0} = 0. \nn
\eea
The last step by virtue of  $G_\emb$ being a subgroup of $G$ implies
that $\Ad(e^{X \theta})\calG_\emb^\perp = \calG_\emb^\perp$.

One should note that the above proof is a modified version of that
given in $\cite{me1}$, to encompass the situation when $\calV_\emb$ is
defined locally. In this context our definition of an embedded defect
is more general than that discussed in \cite{Vach94}, where they
considered $\calV_{\rm emb}$ to be independent of $\theta$ -- which is
equivalent to assuming that $D(e^{X \theta})\Phi_0$ is a circle
centred on the origin within $\calV$. Generically, geodesics on
non-spherical homogenous spaces are not of this form --- an example of
which is given in sec.~(\ref{sec-su5}).

\bigskip
\bigskip

{\noindent{\Large{\bf Appendix B: Inner Product Structures on $\calV$
and $\calG$}}} 
\par
\nopagebreak
\bigskip
\nopagebreak
\noindent
The inner product on $\calV$ is given by the (real) Euclidean inner
product. Considering an element ${\bf v} \in \calV$ as a column vector
the form we shall use is:
\beq
\inprod{{\bf v}_1}{{\bf v}_2} = {\rm Re} [ {\bf v}_1^\dagger {\bf
v}_2 ].
\eeq
This inner product is isometric under automorphisms of $\calV$.

To construct a non-degenerate Ad$(G)$-invariant inner
product on $\calG$, we start by splitting it into its
commuting subalgebras, $\calG_1\oplus\cdots\oplus\calG_n$,
where each $\calG_f$ is either simple or a one-dimensional
$u(1)$ algebra.  For each $\calG_f$ there is a natural
invariant scalar product $\inproda{\cdot}{\cdot}_f$, unique up to a
factor~\cite{oni}.  On a $u(1)$ algebra, we may set $\{X,X\}_f=1$,
where $X$ is the generator normalised so that
$e^{2\pi X}=1$. Upon a simple $\calG_f$ it is (proportional to) the
Killing form; regarding $\calG_f$ as a matrix algebra embedded in
gl$({\bf C}^p)$ for some $p$, the inner product may be defined by 
\beq
\inproda{X}{Y}_f = -p {\rm Re}[{\rm tr}(X^{\dag} Y)].
\eeq
Then the most general Ad$(G)$-invariant scalar product on
$\calG$ is
\beq
\label{scaleip}
\inprod{X}{Y} = \sum_{f=1}^n {s_f}\inproda{X_f}{Y_f}_f,
\eeq
for arbitrary values of the scaling factors $s_f$, where $X =
X_1+\cdots+X_n$, $Y=Y_1+\cdots+Y_n$ and $X_f, Y_f \in \calG_f$. The
scales $s_f$ relate to the norm on each $\calG_f$; for $X_f \in
\calG_f$ such that $\inproda{X_f}{X_f}=1$ the unit norm
generator with respect to $\inprod{\cdot}{\cdot}$ is written $q_f
X_f$, with $q_f$ a function of $\{s_1,\cdots,s_n\}$. 

These two inner products over their respective spaces are related by
the following result:

\result{Theorem}{
The (real) inner product on $\calV$ is related to the inner product 
on $\calM$ by
\bea
\inprod{d(X_i){\bf v}}{d(Y_j){\bf v}} &=& \la_i \la_j
\inprod{X_i}{Y_j}, \ X_i \in \m_i, Y_j \in \m_j,\nn \\
{\rm where}\  \la_i &=&
\frac{\norm{d(X_i){\bf v}}}{\norm{X_i}}. \nn  
\eea
Each $\la_i$ is constant upon its particular $\m_i$.
}\\
The theorem is original to this paper, although it is a fairly simple
application of linear algebra. It is a consequence of two inner
products having a similar invariance, and is most easily derived from:

\result{Lemma}{
Consider $\inprod{\cdot}{\cdot}_1$ and $\inprod{\cdot}{\cdot}_2$ two
(real) inner products on $\calV$. Then $\calV$ decomposes into
$\calV_1 \oplus \cdots \oplus \calV_n$ such that 
\beq
\inprod{{\bf v}_i}{{\bf v}_j}_2 = \la_i \la_j 
\inprod{{\bf v}_i}{{\bf v}_j}_1, \ {\bf v}_i \in \m_i,
{\bf v}_j \in \m_j.
\eeq
Each $\la_i$ is constant over its particular $\calV_i$.
}

\proof{
Choosing a basis for $\calV$ orthonormal with respect to
$\inprod{\cdot}{\cdot}_1$, the inner products can be written
$\inprod{{\bf u}}{{\bf v}}_1={\bf u}^{\top}{\bf v}$ and
$\inprod{{\bf u}}{{\bf v}}_2={\bf u}^{\top} A {\bf v}$, with ${\bf u},
{\bf v} \in \calV$, with $A$ a non-degenerate symmetric matrix. Denote
the eigenspaces of $A$ by $\calV_1 \oplus \cdots \oplus \calV_n$, with
corresponding eigenvalues $\la_i^2$, then for ${\bf v}_i \in \calV_i$
\beq
\inprod{{\bf v}_i}{{\bf v}_j}_2={\bf v}_i^\top A{\bf v}_j=\la_i \la_j
{\bf v}_i^\top {\bf v}_j.
\eeq
Finally it is clear that $\la_i$ is constant over $\calV_i$ because
it is an eigenvalue, but also $\la_i = \norm{{\bf v}_i}_2/\norm{{\bf
v}_i}_1$
} 

Then from this lemma the theorem is easily proved:

\proof{
Firstly, let $\inprod{X}{Y}_1$ denote the usual $\Ad(G)$
invariant inner product over $\calG$, but restricted to $\calM$; note  
that it is $\Ad(H)$ invariant.

Then one may induce a second inner product on $\calM$ from that on
$\calV$: this takes the form
$\inprod{X}{Y}_2 = \inprod{d(X){\bf v}}{d(Y) {\bf v}}$. Again this
inner product is $\Ad(H)$ invariant over $\calM$.

From the lemma $\calM$ reduces to $\widetilde\calM_1 \oplus \cdots
\oplus \widetilde\calM_k$ such that for $X\in\widetilde\calM_i$,
$Y\in\widetilde\calM_j$
\beq
\label{eq}
\inprod{d(X){\bf v}}{d(Y){\bf v}}=\la_i\la_j\inprod{X}{Y}.
\eeq
Here $\la_i=\la(X_i)=\norm{d(X_i){\bf v}}/\norm{X_i}$ is
constant over each $\widetilde\m_i$. However $\la(X)$ is also $\Ad(H)$ 
invariant $\la(\Ad(H)X)=\la(X)$, and thus also constant over each
$\m_i$. Hence each $\widetilde\m_i$ is a direct sum of $\m_i$'s and
Eq.~(\ref{eq}) is true for $X_i \in \m_i$, $Y_j \in \m_j$.  
}

\bigskip
\bigskip

{\noindent{\Large{\bf Appendix C: Dynamical Stability Proofs}}} 
\par
\nopagebreak
\bigskip
\nopagebreak
\noindent
For completeness we give the following proofs to results stated in
section (\ref{dynamical}).

\result{Theorem}{
Vortex generators $X \in {\rm pr}_\calM(\calC)$ such that $X \not\in
\calC$ define embedded vortices that 
are stable in a well defined semi-local limit of the model and are
thus dynamically stable for a region of parameter space around that
semi-local limit. 
}

\proof{
Consider $X \in {\rm pr}_\calM(\calC)$, such that it generates a
closed geodesic. Then $X = X_c + X_h$, with $X_c \in \calC$, such that 
${\rm pr}_\calM (X_c)= X$, generating a $U(1)_c \subseteq C$ and 
$X_h \in \calH$ generating a $U(1)_h \subset H$. These define a
decomposition of the symmetry breaking of the Lie algebras: 
\beq
\calG = \calG' \oplus u(1)_c \rightarrow 
\calH= \calH' \oplus u(1)_h,
\eeq
with $U(1)_c \cap H = 1$. It is now clear that the appropriate
semi-local limit to obtain a symmetry breaking of the form
Eq.~(\ref{eq-22}) is to make coupling constants appertaining to $G'$
vanish. The corresponding vortex generator is $X$. This completes the
proof.} 

\result{Lemma}{
Vortices with a stable semi-local limit are {\em always} generated by
generators in one dimensional $\calM_i$'s.
}

\proof{
The proof that if $\calM_i \subseteq {\rm pr}_\calM(\calC)$ then
$\dim(\calM_i)=1$ relies on the following identification: if
one writes $\calM_{\rm ker}$ as the collection of one dimensional
$\calM_i$'s then necessarily $\calM_{\rm ker} = \{X \in
\calM : {\rm \Ad}(H)X = X\}$. Now we are reduced to showing
${\rm pr}_\calM(\calC) \subseteq \calM_{\rm ker}$.

Consider $(X_c + X_h) \in {\rm pr}_\calM(\calC)$, with $X_c \in \calC$
and $X_h \in \calH$. Then $[\calH, X_c + X_h] \in
\calM$. But also $[\calH, X_c + X_h]=[\calH, X_h] \in \calH$.
Hence $[\calH, X_c +X_h]=0$, and equivalently $\Ad(H)(X_c + X_h)=X_c +
X_h$, proving the result.}

%%%%%%%%%%%%%%%%%%%%%%%%%%%  end of text   %%%%%%%%%%%%%%%%%%%%%%%

%bibliography

%end of document
\end{document}